\documentstyle[12pt]{article}
\input epsf
\begin{document}

\def\question#1{{{\marginpar{\small \sc #1}}}}
\newcommand{\QCD}{{ \rm QCD}^{\prime}}
\newcommand{\MSSM}{{ \rm MSSM}^{\prime}}
\newcommand{\eq}{\begin{equation}}
\newcommand{\en}{\end{equation}}
\newcommand{\bino}{\tilde{b}}
\newcommand{\tsquark}{\tilde{t}}
\newcommand{\gluino}{\tilde{g}}
\newcommand{\photino}{\tilde{\gamma}}
\newcommand{\wino}{\tilde{w}}
\newcommand{\mtilde}{\tilde{m}}
\newcommand{\higgsino}{\tilde{h}}
\newcommand{\gsi}{\,\raisebox{-0.13cm}{$\stackrel{\textstyle>}
{\textstyle\sim}$}\,}
\newcommand{\lsi}{\,\raisebox{-0.13cm}{$\stackrel{\textstyle<}
{\textstyle\sim}$}\,}

\rightline{RU-95-26}
\rightline{hep-ph/9508292}
\rightline{July 11, 1995}
\baselineskip=18pt
\vskip 0.7in
\begin{center}
{\bf \LARGE Phenomenology of Light Gauginos}\\
\vspace{.05in}
{\bf \LARGE II. Experimental Signatures }\\
\vspace*{0.6in}
{\large Glennys R. Farrar}\footnote{Research supported in part by
NSF-PHY-94-23002} \\
\vspace{.1in}
{\it Department of Physics and Astronomy \\ Rutgers University,
Piscataway, NJ 08855, USA}\\
\end{center}
\vspace*{0.2in}
\vskip  0.9in
%\eject
%\vspace*{1.in}

{\bf Abstract:} When SUSY breaking produces only dimension-2 terms in
the minimal supersymmetric standard model, the parameters of the
theory can be rather well constrained.  This paper deals with
strategies for the detection of the new hadrons predicted in
the 1-3 GeV mass range.  Some limits are obtained.  New signatures for
squarks are also given.  Squark masses as small as 45 GeV are not yet
excluded.
\thispagestyle{empty}
\newpage
\addtocounter{page}{-1}
%\tableofcontents
%\listoffigures
\newpage

In a companion Letter\cite{f:101}\footnote{Refered to as (I) below;
notation not defined here can be found there.  A preliminary
discussion of many points developed here and in (I) was given in
\cite{f:99}.}, I outlined the low-energy features of a
particularly economical and attractive form of SUSY-breaking, in which
the coefficients of dimension-3 SUSY breaking operators are
negligible. The absence of these operators is a consequence of a
number of interesting SUSY breaking scenarios and explains the
non-observation of new sources of CP violation generally expected with
SUSY.  Two to four free parameters of the usual
minimal supersymmetric standard model ``MSSM'' ($A$ and the gaugino
masses) vanish at tree level. The allowed range of the remaining SUSY
parameters can be constrained by requiring correct breaking of the
SU(2)$\times$U(1) gauge symmetry and consistency with LEP chargino,
neutralino, and Higgs mass limits. As described in (I), a first
estimate implies $M_0\sim 100-300$ GeV, $\mu \lsi 100$ GeV, and $tan
\beta \lsi 2$. Gauginos are massless at tree level but get calculable
masses through radiative corrections from top-stop and electroweak
(gaugino/higgsino-Higgs/gauge boson) loops. Evaluating these within
the constrained parameter space leads (I) to the predictions: Gluino
mass $m_{\gluino}$ is 100-600 MeV; photino mass $m_{\photino}$ is
$\sim 100-1000$ MeV; lightest chargino has a mass less than $m_W$; the
lightest Higgs boson may be near its present limit; SUSY-breaking
scalar masses can be lighter, and thus electroweak symmetry breaking
more natural, than is now possible in the conventional scenario.

The gluino forms bound states with gluons, other gluinos, and quarks
and antiquarks in a color octet state. The lightest of these, the
spin-1/2 gluon-gluino bound state called $R^0$ was shown ((I) updating
\cite{f:95}) to have a mass $\sim 1.4 - 2.2$ for gluinos in the
100-300 MeV range.  The other lowest lying new hadrons are the
pseudoscalar $\gluino \gluino$ bound state (whose mass should be
approximately the same as the $R^0$ mass), and the lightest
$R$-baryon, the flavor-singlet spin-0 $uds\gluino$ bound state called
$S^0$ whose mass is probably in the range $1 \frac{1}{2} -
2\frac{1}{2}$ GeV.  Higher lying states decay to these via
conventional strong or weak interactions.  In (I) a method was
developed to estimate the lifetime of the $R^0$, which decays to a
photino and hadrons.  For most of the parameter space of interest its
lifetime is longer than about $\sim 10^{-10}$ sec.

Let us begin with discovery strategies for the $R^0$.  The mass and
lifetime estimates given in (I), along with our lack of exact
information about squark masses, leave a large enough range of
uncertainty that for the present we must consider discovery strategies
in two cases: that the $R^0$ can be discovered via its decay, or it is
too long lived for that.  While we should consider detection of
$R$-hadrons for all $\gluino$ and $\photino$ masses in the predicted
ranges, and all $R^0$ lifetimes $\gsi 5 \times 10^{-11}$ sec, the most
interesting portion of the range is that for which the photinos
account for the cold dark matter of the Universe\cite{f:100}: $1.6
\lsi r \equiv m(R^0)/m_{\photino}\lsi 2.2$.  Defining $M_{sq} =
\mu_{sq} \cdot 100$ GeV, the discussion in (I) leads to the lifetime
range $\sim (10^{-10} - 10^{-7})~\mu_{sq}^4$ sec for $r$ in the range
1.6 - 2.2 and $1.4 < M(R^0) < 2$ GeV.  This is comparable to the the
$K_L^0 - K_S^0$ lifetime range if $\mu_{sq} \sim 1$.

In ref. \cite{f:95} I discussed strategies for detecting or excluding
the existance of an $R^0$ with a lifetime so long that it only rarely
decays in the apparatus.  Here I discuss several approaches
appropriate if the $R^0$ lifetime is in the $\sim 10^{-6} - 10^{-10}$s
range.  If $R^0$'s exist, beams for rare $K^0$ decay and
$\epsilon'/\epsilon$ experiments would contain them, and the detectors
designed to observe $K^0$ decays can be used to study $R^0$ decays.
While $R^0$ production cross sections can be reliably computed in
perturbative QCD when the $R^0$'s are produced with $p_{\perp} \gsi 1$
GeV, high-luminosity neutral kaon beams are produced at low
$p_{\perp}$ so pQCD cannot be used to estimate the $R^0$ flux in the
beam.  The most important outstanding phenomenological problem in
studying this scenario is to develop reliable methods for estimating
the $R^0$ production cross section in the low $p_{\perp}$ region; this
problem will be left for the future.  In the remainder of this paper
we will simply paramterize the ratio of $R^0$ to $K_L^0$ fluxes in the
beam at the production target by $p \cdot 10^{-4}$.

The momentum in the $R^0$ rest frame of a hadron $h$, produced in the
two body decay $R^0 \rightarrow \photino +~ h$, is:
\eq
P_h = \frac{\sqrt{m_R^4 + m_{\photino}^4 + m_h^4 - 2 m_R^2
m_{\photino}^2 - 2 m_{\photino}^2 m_h^2 - 2 m_h^2 m_R^2}}{2 m_R}.
\label{P}
\en
This falls in the range 350-800 MeV when $h= \pi^0$, for the mass
ranges of greatest interest: $1.2 ~{\rm GeV} < m_{R^0} < 2 ~{\rm GeV}$
and $ 1.6 < r ~\lsi ~2.2$.  Therefore, unless the $R^0$ is in the
extreme high end of its mass range and the photino is in the low end
of its estimated mass range, final states with more than one hadron
will be significantly suppressed by phase space\footnote{For instance
the final state $\pi^+\pi^-\pi^+\pi^-\photino$ suggested by Carlson
and Sher, while certainly distinctive, has a very small branching
ratio for practically all the masses under consideration.}.

A particularly interesting decay to consider is $R^0 \rightarrow \eta
\photino$.\footnote{I thank W. Willis for this suggestion.}  Since
$m(\eta) = 547~ {\rm MeV} > m(K^0) = 498$ MeV, there would be very
little background mimicking $\eta$'s in a precision $K^0$-decay
experiment, so that detecting $\eta$'s in the decay region of one of
these experiments (e.g., via their $\pi^+ \pi^- \pi^0$ final state
whose branching fraction is 0.23) would be strong circumstantial
evidence for an $R^0$.  Since the $R^0$ is a flavor singlet and the
$\photino$ is a definite superposition of isosinglet and isovector,
the relative strength of the $R^0 \rightarrow \pi^0 \photino$ and $R^0
\rightarrow \eta \photino$ matrix elements is determined by Clebsches
and is $3:1$.  Thus the branching fraction of the $\eta \photino$
decay mode is about 10\%, in the most favorable case that multibody
decay modes and phase space suppression of the $\eta$ relative to the
$\pi^0$ are unimportant.  If $\eta$'s are detected, the Jacobian peak
in the $\eta$ transverse momentum, which occurs at $p_{\perp} \approx
P_{\eta}$ defined in eq. (\ref{P}) above, gives both a confirming
signature of its origin, and provides information on the allowed
regions of $R^0$ and $\photino$ masses.

\begin{figure}
\epsfxsize=\hsize
\epsffile{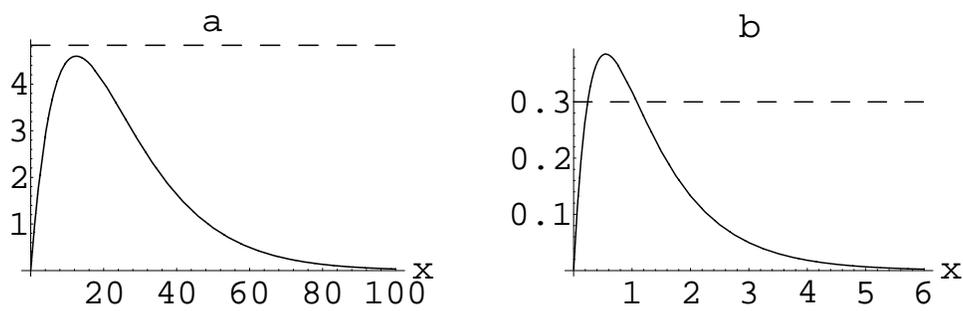}

\caption{Sensitivity of (a) $K_L^0$ beam with $x^{lim} = 4.8$ and (b)
an NA48-like $K_S^0$ beam, with $x^{lim} = 0.3$.  }
\label{kbeams}
\end{figure}

We can estimate the sensitivity of the next generation of neutral kaon
experiments to $R^0$'s as follows.  The number of decays of a particle
with decay length $\lambda \equiv <\gamma \beta c \tau>$, in a
fiducial region extending from $L$ to $L+l$, is
\eq
N = N_0 \left( e^{-\frac{L}{\lambda}} - e^{-\frac{L+l}{\lambda}}
\right),
\label{decays}
\en
where $N_0$ is the total number of particles leaving the production
point.  In the $\frac{\epsilon'}{\epsilon}$ experiments which will
begin running during 1996 at FNAL and CERN (KTeV and NA48), $L\sim
120$ m, $l \sim 12-30$ m, and $\lambda_{K_L^0} \sim 4.5$ km, so $
e^{-\frac{L}{\lambda}} - e^{-\frac{(L+l)}{\lambda}}
\approx \frac{l}{\lambda}e^{-\frac{L}{\lambda}}$.  Denote by
$N_{R^{+-0}}$ the number of reconstructed $R^0 \rightarrow \pi^+ \pi^-
\pi^0 \photino$ events, in which the $\pi^+ \pi^- \pi^0$ invariant
mass is that of an $\eta$, and by $N_{K_L^{00}}$ the number of
reconstructed $K_L \rightarrow \pi^0 \pi^0$ events.  Then defining
$br(R^0 \rightarrow \eta \photino) \times br(\eta \rightarrow \pi^+
\pi^- \pi^0) \equiv b~ 10^{-2}$, using $br(K_L \rightarrow \pi^0
\pi^0) = 9 ~10^{-4}$, and idealizing the particles as having a narrow
energy spread, eq. (\ref{decays}) leads to:
\eq
N_{R^{+-0}} \approx N_{K_L^{00}} (~p~10^{-4})
\left(\frac{b~10^{-2}}{9~10^{-4}}\right)
\left(\frac{\epsilon^{+-0}}{\epsilon^{00}} \right)
\frac{<\gamma \beta \tau>_{K^0_L}}{<\gamma \beta \tau>_{R^0}}
exp[ -L/<\gamma \beta c \tau>_{R^0}],
\label{etas}
\en
where $\epsilon^{+-0}$ and $\epsilon^{00}$ are the efficiencies for
reconstructing the $\pi^+ \pi^- \pi^0$ final state of the $\eta$ and
the $\pi^0 \pi^0$ final state of the $K_L^0$ respectively, $\gamma =
\frac{E}{m}$ is the relativistic time dilation factor, and $\beta =
\frac{P}{E}$ will be taken to be 1 below.  Introducing $x \equiv
\frac{<E_{K_L^0}>m_{R^0}\tau_{K_L^0}}{<E_{R^0}>m_{K^0_L}\tau_{R^0}}$,
we have
\eq
x~ exp[-\frac{L}{\lambda_{K^0_L}}x] = \frac{0.9
{}~10^3}{pb}\frac{N_{R^{+-0}}}{N_{K_L^{00}}}
\frac{\epsilon^{00}}{\epsilon^{+-0}}\equiv x_{\eta (L)}^{lim}.
\label{xrangeeta}
\en
For these experiments $\frac{L}{\lambda_{K_L^0}} \approx 0.08$, so if
$\frac{\epsilon^{00}}{\epsilon^{+-0}pb} \sim 10$ and we demand three
reconstructed $\eta$'s so $N_{R^{+-0}} = 3$, a sensitivity $x_{\eta
(L)}^{lim} = 5.4~10^{-3}$ is reached after collecting $5~10^{6}$
reconstructed $ K_L \rightarrow \pi^0 \pi^0$ events, typical of the
next generation of $\frac{\epsilon'}{\epsilon}$ experiments.  Such a
sensitivity allows the range $0.0054 < x < 103$ to be probed.  This
translates to an ability to discover $R^0$'s with a lifetime in the
range $\sim 2~ 10^{-9}- 4~ 10^{-5}$ sec, using $x = 4
\frac{\tau_{K_L^0}}{\tau_{R^0}}$.  For shorter lifetimes, the
$R^0$'s decay before reaching the fiducial region, while for longer
lifetimes not even one reconstructed $R^0$ decay is expected during
the experiment.  The dependence of the lifetime reach on the
efficiency and production rates, whose combined effect is contained in
$x_{\eta (L)}^{lim}$, is illustrated in Fig. \ref{kbeams}a showing the
left and right hand sides of eq. (\ref{xrangeeta}) for $x_{\eta
(L)}^{lim} = 4.8$, of interest below. As one would expect, the reach
to longer lifetimes (small $x$) is extremely sensitive to the event
rate while the short lifetime cutoff has a relatively small variation
as $x_{\eta (L)}^{lim}$ varies.  Note that in a rare $K_L^0$-decay
experiment the flux of $K_L^0$'s is much greater than for the
$\frac{\epsilon'}{\epsilon}$ experiments, so a greater sensitivity can
be achieved for a comparable acceptance.  Unfortunately, the E799
trigger did not accept such events.

Use of an intense $K_S^0$ beam would allow shorter lifetimes to be
probed. The FNAL E661 experiment designed to search for the CP
violating $K_S^0 \rightarrow \pi^+ \pi^- \pi^0$ decay had a high
$K_S^0$ flux and a decay region close to the production
target.  However its 20 MeV invariant mass resolution may be
insufficient to adequately distinguish $\eta$'s from $K^0$'s.
Unfortunately, the $K^0_S$ flux planned for upcoming experiments is
inadequate to improve upon the limits which will be obtained from the
$K^0_L$ beams.  We can repeat the analysis above for the NA48
$\frac{\epsilon'}{\epsilon}$ experiment, taking into account that
for their $K_S^0$ beam $\lambda_{K_S^0} \approx L \approx l/2$.  In
this case $x$ must satisfy
\eq
\left( e^{-\frac{Lx}{\lambda_{K_S^0}}} -
e^{-\frac{(L+l)x}{\lambda_{K_S^0}}} \right) <
\left( e^{-\frac{L}{\lambda_{K_S^0}}} -
e^{-\frac{(L+l)x-}{\lambda_{K_S^0}}} \right)\frac{br(K_S^0 \rightarrow
\pi^0 \pi^0)}{b ~10^{-2} p~10^{-4}}\frac{N_{R^{+-0}}}{N_{K_L^{00}}}
\frac{\epsilon^{00}}{\epsilon^{+-0}}\equiv x_{\eta (S)}^{lim}.
\en
Taking the same production rate and efficiencies as before and
assuming $\sim 10^7$ reconstructed $K_S^0 \rightarrow \pi^0 \pi^0$
decays gives $x_{\eta (S)}^{lim} =  0.3$.  The left and right hand sides of
this equation is shown in Fig. \ref{kbeams}b for this value of
$x_{\eta (S)}^{lim}$.  The sensitivity range $0.24 < x < 1.1$ is much less
than in the $K_L^0$ beams, simply because their $K_S^0$ beam is roughly
three orders of magnitude lower in intensity than their $K_L^0$ beam.

\begin{figure}
\epsfxsize=\hsize
\epsffile{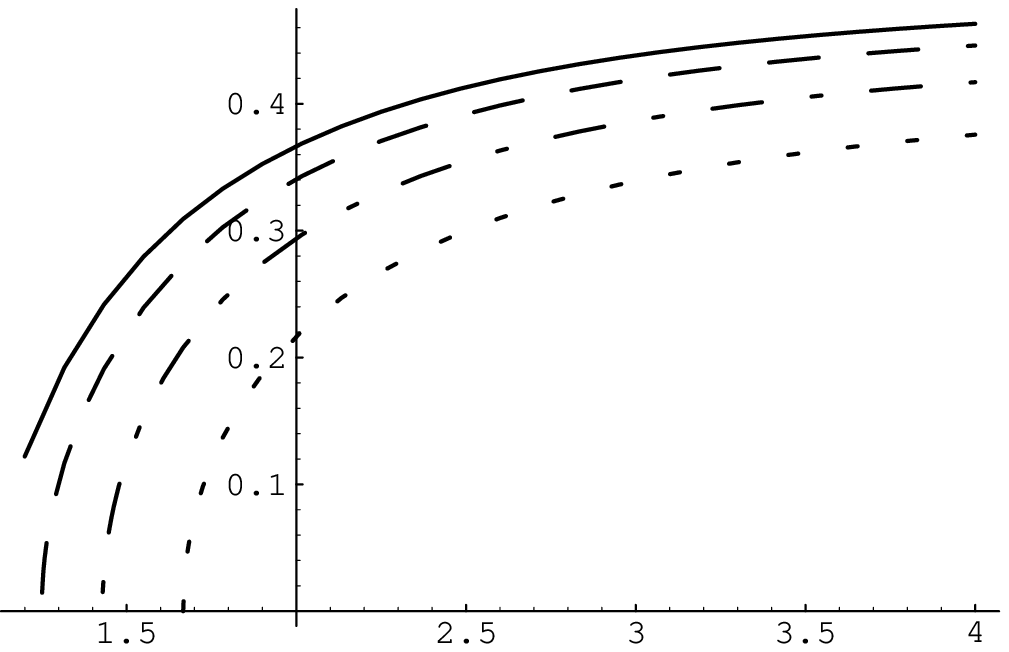}

\caption{$P_h$ in units of $m(R^0)$ as function of $r\equiv
\frac{m(R^0)}{m_{\photino}}$, for $\frac{m_h}{m(R^0)} = 0.1$ (solid),
0.2 (dashed), 0.3 (dot-dashed), and 0.4 (dotted).}
\label{P_h}
\end{figure}

The possibility that photinos account for the cold dark matter of the
universe leads us to be particularly interested in masses for which $r
= \frac{m_R}{m_{\photino}}  \lsi 2.2$\cite{f:101}.  Since the phase space
volume $\sim P_h^2$, we see from Fig. \ref{P_h} which shows how $P_h$
depends on $r$, that in the $r$ region of interest the $R^0 \rightarrow
\photino \eta$ decay may be considerably kinematically suppressed
compared to $R^0 \rightarrow \photino \pi^0$. For instance for $r =
1.6$ and $M_{R^0} = 1.7$ GeV, the branching fraction for $R^0
\rightarrow \photino \pi^0$ should be about 97\%, while the branching
fraction for $R^0 \rightarrow \photino \eta$ is about 3\% and drops
rapidly for smaller $M_{R^0}$.  Therefore it would be very attractive
to be able to identify the $\pi^0$ plus missing photino final state in
the $K^0$ beam experiments.  This is demanding technically, but
justifies the effort.  Even though the overall kinematics of
individual decays is unknown, both $m_{\photino}$ and $M_{R^0}$ can be
determined if $p_{\perp}^{max}$ is measured for {\it both} $\pi$ and
$\eta$ final states, because eq. (\ref{P}) gives two conditions fixing
the two unknowns, $m(R^0)$ and $m_{\photino}$, in terms of the
observables, $P_{\pi}$ and $P_{\eta}$.  When the photino mass, the
$R^0$ mass and lifetime, and the cross section for $R^0 ~N \rightarrow
\photino ~X$ have been measured\footnote{For cosmology one actually
needs $\sigma(R^0 \pi \rightarrow \photino \pi) $ but  that can be far
better estimated when $\sigma(R^0 N\rightarrow \photino N) $ is
known.} it will be possible to refine the estimate of the critical
value of $r$.  This will permit confirmation or refutation of the
proposal\cite{f:100} that relic photinos are responsible for the bulk
of the missing matter of the Universe.

The Fermilab E799 experiment obtained\cite{ktev:pi0nunubar} a 90\% cl
limit $br(K_L^0 \rightarrow \pi^0 \nu \bar{\nu}) \lsi ~5.8 ~10^{-5}$,
which can already be used to limit the $R^0$ lifetime if the $R^0$ flux
is $\gsi 10^{-4}$ of the $K^0_L$ flux.  In that experiment, pions were
required to have transverse momentum in the range $160 < P_t < 231$
MeV/c. For a given flux of $K_L^0$'s, the ratio of the number of
$\pi^0$'s in this $P_t$ range coming from $R^0 \rightarrow \pi^0
\photino$ compared to those coming from $K_L^0 \rightarrow \pi^0 \nu
\bar{\nu}$ is
\eq
 ~p~10^{-4} \frac{br(R^0 \rightarrow \pi^0
\photino)}{br(K^0_L \rightarrow \pi^0 \nu \bar{\nu})}
\frac{\epsilon^{0}}{\epsilon^{0 \nu \bar{\nu} } }
\frac{<\gamma \beta \tau>_{K^0_L}}{<\gamma \beta \tau>_{R^0}}
exp[ -L/<\gamma \beta c \tau>_{R^0}],\nonumber
\label{pi0s}
\en
where $\epsilon^{0}$ is the fraction of $\pi^0$'s in $R^0 \rightarrow
\pi^0 \photino$ having $160 < P_t < 231$ MeV/c times the $\pi^0$
detection efficiency, $\epsilon^{0 \nu \bar{\nu}}$ is the same thing for
$\pi^0$'s coming from $K_L^0 \rightarrow \pi^0 \nu \bar{\nu}$, and we
neglect depletion of the $K^0_L$ beam by decays before the fiducial
region.  Since the $\pi^0$ detection efficiency is the same in the two
cases, $\frac{\epsilon^{0}}{\epsilon^{0 \nu \bar{\nu} } } $ is just
the ratio of probabilities (which we will denote respectively
$f_K$ and $f_R$) for the $\pi^0$ to have $160 < P_t < 231$ GeV in the
two cases.  Thus in terms of the variable $x$ used previously, this
excludes the region for which
\eq
x Exp[-\frac{L}{\lambda_{K^0_L}}x] \ge \frac{5.8 ~10^{-5}~f_K}{p
{}~10^{-4}~ f_R } \equiv x_{\pi (L)}^{lim}.
\label{xrangepi}
\en
With the spectrum $\frac{d\Gamma}{d E_{\pi^0}}$ used in ref.
\cite{ktev:pi0nunubar}, $f_K = 0.5$.  For $R^0 \rightarrow \pi^0
\photino$, $f_R = \frac{(231)^2 - (160)^2}{P_{\pi}^2} \sim (0.06 -
0.11)$, when $M_{R^0} = 1.7$ GeV and $r$ is in the range $2.2 - 1.6$.
Taking the smallest of these values ($f_R = 0.06$) to be conservative
gives $x_{\pi (L)}^{lim} =  4.8/p$ so that for $p=1$ no limit is
obtained, as can be seen from Fig. \ref{kbeams}a.  If all pions with
$P_t>160$ MeV/c are retained in the analysis, 6 rather than 0 events
pass the cuts\footnote{E799, private communication.}.  With this
larger $P_t$ acceptance, $f_R$ increases to $0.94$, so the sensitivity
improves to $x_{\pi (L)}^{lim} = 1.4/p$. I have used the fact that
seeing no events, a 90 \% cl limit is calculated as if there are 2.3
events, while with 6 events it is calculated as if there are 10.5
events\cite{pdg94}.   Now the range $1.6 < x < 42 $ is excluded for
$p=1$.  Again assuming $x \sim 4 \frac{\tau_{K_L^0}}{\tau_{R^0}}$,
this excludes the lifetime range $\sim (5~ 10^{-9} -  10^{-7})$ sec.
However, the entire lifetime range is allowed if $p \lsi 0.3$.  Thus
it is clear that a good understanding of the expected production cross
sections is necessary before one can set limits on the allowed $R^0$
lifetime.  In conclusion, this discussion shows that the decay $R^0
\rightarrow \pi^0 \photino$ is feasible to study experimentally.  The
increase in branching ratio in comparison to the $\eta \photino$ final
state is capable of offsetting the cuts needed to reduce background to
the solitary $\pi^0$ final state.  Even though the existing
limit\cite{ktev:pi0nunubar} is inadequate to definitively exclude any
part of the $R^0$ lifetime range, on account of the present theoretical
uncertainty about the $R^0$ production parameter $p$, the space of
allowed lifetime vs production cross section has been restricted,
which will be useful for planning future experiments.

Before leaving the topic of detecting $R^0$'s decaying in
a $K_L^0$ beam, we estimate the branching fraction for the
reaction $br(R^0 \rightarrow \pi^+ \pi^- \photino)$.  While not as
dramatic as detecting $\eta$'s, a $\pi^+ \pi^-$ pair with several
hundred MeV of transverse momentum and invariant mass greater than
$m_K$ would nonetheless be a rather background-free signal.  For
some experimental setups, this is a more tractable final state to
reconstruct than one including a $\pi^0$.  As noted above, the
coupling of $R^0$ to photino plus isotriplet is three times the
coupling to the same particles in an isosinglet state.  Thus the 2
pions in $R^0 \rightarrow \pi \pi \photino$ are 90\% of the time in an
$I=1,~I_z=0$ state which is necessarily $\pi^+ \pi^-$, and 10\% of the
time in an $I=0$ state which is $2/3$ of the time $\pi^+ \pi^-$.
Therefore 97\% of the $\pi \pi \photino$ final state will be in the
favorable $\pi^+ \pi^- \photino$ channel.  Since the $\pi^0 \photino$
final state completely dominates the decay rate, we have $br(R^0
\rightarrow \pi^+ \pi^- \photino) \sim \Phi_3/\Phi_2/\Lambda^2$, where
$\Phi_n$ is the phase space for the $n$-particle final state and
$\Lambda$ is some characteristic mass scale of the problem.  For
$M_{R^0}=1.7~(1.4)$ GeV and $r$=2, taking $\Lambda = m_{\rho}$ gives
$br(R^0 \rightarrow \pi^+ \pi^- \photino) \sim 6~10^{-3}~(3~10^{-3})$.
It seems unlikely that the characteristic mass scale of the problem
would be as low as 200 MeV, but if it were, the branching fraction
estimate should be increased by a factor of $\sim 15$.  The loss due
to requiring $M(\pi^+ \pi^-) > M_K$ is not severe: e.g., for $M_{R^0}
= 1.7$ GeV and $r=2$, 72\% of the events would pass this cut.

Given $br(R^0 \rightarrow \pi^+ \pi^- \photino)$ we can evaluate the
sensitivity of Bernstein et al\cite{bernstein} to $R^0$'s.  This
experiment placed limits on the production cross section times
branching ratio of a neutral hadron decaying into charged
particles, with lifetimes in the range $10^{-8} - 2 \times 10^{-6}$
sec and masses between 1.5 and 7.5 GeV, by looking for a deviation
from a smooth decrease in the transverse momentum distribution. The
analysis assumed that the final state particles were all
pions\footnote{G. Thomson, private communication.}, so that the
Jacobian peak in their transverse momentum falls at approximately half the
mass of the decaying particle.  When $m_{\photino} \sim \frac{1}{2}
m(R^0)$ however, the peak in the transverse momentum of the pions
falls at a much lower value: about 350 MeV for an $R^0$ mass of 1.5
GeV, where their cross section limit is most stringent.  Since the
background in that transverse momentum bin is a factor of $\sim 100$
larger than at 750 MeV, the sensitivity is reduced by at least a
factor of $\sim 10$.  Combining our estimate for $br(R^0 \rightarrow
\pi^+ \pi^- \photino) \sim 3~10^{-3}$ with this reduction in
sensitivity and the 1.5 reduction in sensitivity due to having a
3-body final state leads to a limit on the $R^0$ production cross
section in 400 GeV $p + Be$ collisions at $x_f = 0.2$ and $p_{\perp} =
0$ of $\frac{E d  \sigma}{d^3  p} \lsi 2.5 10^{-30} cm^2/GeV^2$ or
higher, in their most sensitive lifetime region of $\tau = 3~ 10^{-8}$
sec.  This is about a factor of 16 lower than the production cross
section for $\bar{\Xi}^0$.  For comparison, the $\bar{\Xi}^0$
invariant cross section is a factor of 25 smaller than that of the
$\bar{\Lambda}^0$ in the same kinematic region, while
$m(\bar{\Xi}^0)/m(\bar{\Lambda}^0) \approx 1.5 {\rm
GeV}/m(\bar{\Xi}^0)$.  The experiment does not give limits for lower
$R^0$ mass while for larger mass the limits are worse than this.  Thus
I conclude that the Bernstein et al experiment would not be expected
to have seen $R^0$'s.

There is another interesting ground-state $R$-hadron besides the $R^0$,
namely the flavor singlet scalar baryon $uds \gluino$ denoted $S^0$.
On account of the very strong hyperfine attraction among the quarks in
the flavor-singlet channel\cite{f:52}, its mass is about $210 \pm
20$ MeV lower than that of the lowest R-nucleons.  It is even possible
that the $S^0$ might be close in mass to the $R^0$.  If the baryon
resonance known as the $\Lambda(1405)$, whose properties have not been
easy to understand within conventional QCD, is a ``cryptoexotic''
flavor singlet bound state of $udsg$ as suggested in \cite{f:99}, one
would expect the corresponding state with gluon replaced by a light
gluino to be similar in mass.  In any case, the mass of the $S^0$ is
surely less than $m(\Lambda) + m(R^0)$, so it does not decay through
strong interactions.  Its mass is also expected to be less than $m(p)
+ m(R^0)$, so there must be a photino rather than $R^0$ in the final
state of its decay.  Therefore the $S^0$ has an extremely long lifetime since
its decay requires a flavor-changing-neutral weak transition as well
as an electromagnetic coupling, and is suppressed by $M_{sq}^{-4}$.
It could even be stable, if $m(S^0) - m(p) - m(e^-) <
m_{\photino}$ and $R$-parity is a good quantum number.  This is not
experimentally excluded\cite{f:51,f:95} as long as the $S^0$
does not bind to nuclei\footnote{Even if R-parity is violated so the
photino decays (e.g.,$\photino \rightarrow \nu \gamma$) the $S^0$
could nevertheless be stable if R-parity is only violated in
conjunction with lepton number violation and not baryon number
violation\cite{masiero_valle}.  Stable relic $S^0$'s could make
up part of the missing mass in our galaxy, but might be too
dissipative to make up the bulk of the cold dark matter and still be
consistent with the observed spectrum of density perturbations.  Since
they are neutral and cannot form nuclei (I), they would not form
atoms.  Thus their energy dissipation would be entirely through the
much lower cross section strong interaction leading them to clump
less than ordinary matter, but still much more than conventional
WIMPS.  It is not evident that the relic density of $S^0$'s
would give the correct amount of dark matter, for plausible values of
the $S^0$ mass, as the photino does\cite{f:100}.  Since the $S^0$ is
not charged, its scattering from photons in the cosmic microwave
background radiation is much smaller than for protons.  Therefore
$S^0$'s could be responsible for the very high energy cosmic ray
events recently observed without being required to originate
uncomfortably close to our galaxy as required for
protons\cite{greisen_zatsepin}.}.  There is not a first-principles
understanding of the intermediate-range nuclear force, so that it is
not possible to decide with certainty whether the $S^0$ will bind to
nuclei.  However the two-pion-exchange force, which is attractive
between nucleons, is repulsive in this case\cite{f:95} because the
mass of the intermediate $R_{\Lambda}$ or $R_{\Sigma}$ is much larger
than that of the $S^0$.  For further discussion of the $S^0$ and other
$R$-hadrons see refs. \cite{f:51}, \cite{f:95}, and (II).

The $S^0$ can be produced via a reaction such as  $
K ~ p \rightarrow R^0 ~ S^0 + X$, or can be produced via decay of a
higher mass $R$-baryon such as an $R$-proton produced in $
p ~ p \rightarrow R_p ~ R_p + X$.  In an intense proton beam at
relatively low energy, the latter reaction is likely to be the most
efficient mechanism for producing $S^0$'s, as it minimizes the
production of ``extra'' mass.  One strategy for finding evidence for
the $S^0$ would be to perform an experiment like that of Gustafson et
al\cite{gustafson}, in which a neutral particle's velocity is measured
by time of flight and its kinetic energy is measured in a calorimeter.
This allows its mass to be determined via the relation $KE =
m(\frac{1}{\sqrt{1- \beta^2}} -1)$.  On account of limitations in time
of flight resolution and kinetic energy measurement, ref.
\cite{gustafson} was only able to study masses $> 2$ GeV, below which
the background from neutrons became too large.  An interesting aspect
of using a primary proton beam at the Brookhaven AGS, where the
available cm energy is limited ($p_{beam} \sim 20$ GeV), is that
production of pairs of $S^0$'s probably dominates associated
production of $S^0$-$R^0$ or production of $R^0$ pairs, due to the
efficiency from an energy standpoint of packaging baryon number and
R-parity together in an $S^0$ or $R_p$ whose mass is probably much
less than the combined mass of a nucleon and an $R^0$. This should
give an extra constraint which can help discriminate against the
neutron background in such a search.  Likewise, a low energy $S^0$
will typically remain an $S^0$ while scattering in matter, rather than
convert to an $R^0$ via, e.g., $S^0 ~ N \rightarrow R^0 ~\Lambda ~ N'
+ X$, because the mass of the $R^0 ~\Lambda$ system is of order 1 GeV
larger than that of the $S^0$. This assures that for sufficiently low
energy $S^0$'s, the calorimetric determination of the $S^0$ kinetic
energy is not smeared by conversion to $R^0$.  Although the $S^0$ has
approximately neutron-like interaction with matter, its cross section
could easily differ by a factor of two or more, so that the systematic
effects on the calorimetry of the unknown $S^0$ cross section must be
studied.  Fortunately the behavior of a neutron in a calorimeter can
be independently determined by virtue of its similarity to a proton
for which there can be no issue of contamination by $S^0$'s in a test
beam.

If the $R^0$ is too long-lived to be found via anomalous decays
in kaon beams and the $S^0$ cannot be discriminated from a neutron,
a dedicated experiment studying two-body reactions of the type
$R^0 + N \rightarrow K^{+,0}+ S^0$ could be done.  Depending on the
distance from the primary target and the nature of the detector, the
backgrounds would be processes such as $K^0_L + N \rightarrow K^{+,0} +
n$, etc.  If the final state neutral baryon is required to rescatter,
and the momentum of the kaon is determined, and time of flight is used
to determine $\beta$ for the incident particle, all with sufficient
accuracy, one would have enough constraints to establish that one was
dealing with a two-body scattering and to determine the $S^0$ and
$R^0$ masses.  Measuring the final neutral baryon's kinetic energy
would give an over-constrained fit which would be helpful.

Light $R$-hadrons other than the $R^0$ and $S^0$ will decay, most via
the strong interactions, into one of these.  However since the
lightest $R$-nucleons are only about $210 \pm 20$ MeV heavier than the
$S^0$, they would decay weakly, mainly to $S^0 \pi$.  Any model
which correctly accounts for the regularities of hyperon lifetimes and
branching fractions should be able to give a reliable estimate for the
$R$-nucleon lifetimes.  We can expect them to be in the range $\sim 2~
10^{-10}-10^{-11}$ sec, since the $Q$ value of the decay is comparable
to those of the hyperons, whose lifetimes are around $10^{-10}$ sec,
while simply scaling down the average $\Sigma^{\pm}$ lifetime by the
fifth power of the mass of the decaying
particle\footnote{Dimensionally $\tau \sim M_{sq}^4/M^5$, so this
would be the correct procedure if the $Q$-value scaled as well.} gives
$\tau(R_N) \sim 1.2 \times 10^{-11} {\rm sec} [(m(R_N)/(1.8 ~{\rm
GeV})]^{-5}$. As can be seen from \cite{f:95}, existing experimental
limits do not apply to the lifetime region of interest.  Silicon
microstrip detectors for charm studies are unlikely to be very useful
since they are optimized for the lifetime range $(0.2 - 1.0)~10^{-12}$
sec.  Unlike ordinary hyperon decay, no more than one final state
particle is charged, except for very low
branching fraction reactions such as $R_n \rightarrow S^0 ~\pi^-~ e^+
{}~\nu_e$, or $R_n \rightarrow S^0~\pi^0$ followed by $\pi^0 \rightarrow
\gamma~ e^+~ e^-$.  In order to distinguish the decay from the much
more abundant background such as $\Sigma^+ \rightarrow n ~ \pi^+$,
which has a very similar energy release, one could rescatter the final
neutral in order to get its direction.  Then with sufficiently
accurate knowledge of the momentum of the initial charged beam and the
momentum (and identity) of the final pion, one has enough constraints
to determine the masses of the initial and final baryons.  The
feasibility of such an experiment is worth investigating.  One other
charged $R$-baryon could be strong-interaction stable, the
$R_{\Omega^-}$.  Assuming its mass is 940 MeV ($= m(\Omega^-)-m(N) +
210$ MeV) greater than the $S^0$ mass, one would expect it to decay
weakly to $R_{\Xi} + \pi$ or $R_{\Sigma}+ K$, with the $R_{\Xi}$ or
$R_{\Sigma}$ decaying strongly to $S^0 K$ or $S^0 \pi$ respectively.
This would produce a more distinctive signature than the $R$-nucleon
decays, but at the expense of the lower production cross section for
$R_{\Omega^-}$'s than for R-nucleons.

In addition to the new hadrons expected when there are light gluinos
in the theory, there are also many other consequences of light
gluinos.  None of them are presently capable of settling the question
as to whether light gluinos exist, since they all rely on
understanding non-perturbative aspects of QCD.  Existing models of
non-perturbative behavior are {\it tuned} to agree with data assuming
the validity of standard QCD without gluinos.  Adding gluinos to the
theory is practically certain to cause a deterioration of the fits.
Nonetheless it is interesting to recall that jet production at LEP and
FNAL should be different with and without light gluinos\cite{f:82}.
Since gluinos in this scenario live long enough  that they hadronize
before decaying to a photino, they produce jets similar to those
produced by the other light, colored quanta: gluons and quarks.  In
$Z^0$ decay, only 4- and more- jet events are modified; the magnitude
of the expected change is smaller than the uncertainty in the
theoretical prediction\cite{f:82}. Calculation of the 1-loop
corrections to the 4-jet amplitudes would allow the theoretical
uncertainty to be reduced sufficiently that data might be able to
discriminate between QCD with and without gluinos\cite{f:82}.  In $p
\bar{p}$ collisions, there is a difference already in 1-jet cross
sections. However absolute predictions are more difficult than for
$Z^0$ decay since they rely on structure functions which have so far
been determined assuming QCD without gluinos.   More promising might
be to search for differences in the expected {\it relative} $n$-jets
cross sections\cite{f:82}.

Now let us move on to means of detecting evidence of new particles
other than gluinos.  Conventional squark limits do not apply when the
gluino is light and long-enough-lived to hadronize, as in this scenario.
Squarks will be produced in pairs at colliders such as the Tevatron,
and decay immediately, generally to a gluino and a quark.  The gluino
and quark produce jets, so that squark pairs will lead to events with
at least four jets  in which TWO pairs of jets reconstruct to an
invariant mass peak.  To the extent that splitting between mass
eigenstates of each flavor of squark can be neglected, both pairs of
jets should reconstruct to the same mass.  Furthermore, it is
reasonable to expect that the squarks associated with the $u,~d,~s,~c$
and $b$ quarks will be approximately degenerate, while the stop will be
significantly heavier.  If these approximations are better than
the experimental resolution, 5/6 of the signal (all but the stop
pairs) will contribute at the same value of invariant mass.  The cross
section for producing squark pairs is the same as in the conventional
picture, and roughly speaking is about half that of producing a $t
\bar{t}$ pair of the same mass, for each flavor, so there should be a
substantial number of events containing squark pairs at FNAL, up to
quite high squark mass.  A search for events in which {\it two} pairs
of jets reconstruct to the same invariant mass should be made.
Hopefully the experimental dijet-invariant-mass resolution is good
enough that the QCD background will not overwhelm the signal.  For
large enough squark mass the best channel to study is associated
production of squark and gluino, either at $O(\alpha_s^2)$ via
quark-gluon fusion or at $O(\alpha_s^3)$ via gluon fusion.
Squark+gluino final states have three jets, two of which reconstruct
to a definite invariant mass.  Since this signal is less distinctive
than that of squark pairs, QCD background is likely to be a greater
problem.

Squarks generally decay to gluino and quark, but a squark also
decays to a photino and quark with a branching fraction $Q_{sq}^2
\alpha_{em}/(4/3 \alpha_s)$. For a charge +2/3 squark this occurs about
2\% of the time. To find these events, a trigger on missing energy
accompanied by three or more jets can be used. Then a peak should
appear in the invariant mass of one pair of the jets.  Missing
energy alone is a much less efficient tool for finding squarks than in
the conventional scenario.  Furthermore the search reported in ref.
\cite{cdf:gluinolim2} required that the leading jet have NO other jet
opposite in $\phi$ which roughly speaking, would reject 2/3 of the real
events.  Averaging over the $u,~d,~s,~c$ and $b$ quarks, and taking
into account the $\phi$ cut, leads to a factor $\sim 200$ reduction in
the number of events with missing energy compared to the case that
both squarks always decay to a photino and quark jet.  The missing
$E_T$ spectrum is softer as well.

While awaiting a reanalysis of the Tevatron collider data, our best
limit on squark masses is obtained by requiring the squarks not add
too much to the $Z^0$ hadronic width.  The expected change in the
hadronic width of the $Z^0$ is 21.3 times the contribution to the
width of the $Z^0$ from a selectron of the same mass, assuming the
$u,~d,~s,~c$ and $b$ squarks are degenergate and using width ratios
given in \cite{hk:susy}.  Thus the limit on ``extra'' hadronic width
of the $Z^0$, $\Gamma_X < 46$ MeV\cite{aleph:survey}, requires the
squarks to be heavy enough that $\frac{21.3}{4}\beta^3 < 0.27$.  This
implies that if there are five degenerate ``light'' squarks, their
mass must be greater than 42 GeV.  If only one parity eigenstate of a
single flavor of squark is light, this limit is reduced to 27.5 GeV.
Note than any excess width would be entirely in 4 jet events, which
might allow the limit from LEP to be improved, but clearly only
slightly in the case where it is already 42 GeV.

In summary, we have seen that the phenomenology of SUSY breaking
without dimension-3 operators is very rich and accessible.  Present
limits are shockingly weak in comparison to the usual scenario.

{\bf Acknowledgements:}  I am indebted to many people for information
and helpful discussions and suggestions including M. Calvetti, J. Conway,
T. Devlin, J. Kane, L. Littenberg, I. Mannelli, M. Schwartz, S. Somalwar, G.
Thomson, W. Willis, B. Winstein, and M. Witherell.

%\appendix

%\newpage

%\bibliography{susy,qcd,f,radecay,cosmo}
%\bibliographystyle{unsrt}

\end{document}